\documentclass[11pt]{article}
\usepackage{graphicx}
\usepackage{moriond}
\usepackage{amsmath}
\usepackage{amssymb}

\newcommand{\Photo}{}

\newcommand{\cR}{\mathcal{R}}

\begin{document}
\vspace*{4cm}
\title{THE ELECTRON EDM AND EDMS IN TWO-HIGGS-DOUBLET MODELS}

\author{ MARTIN JUNG }

\address{Institut f\" ur Physik, Technische Universit\"at Dortmund\\ D-44221 Dortmund, Germany}

\maketitle\abstracts{
Electric dipole moments constitute highly sensitive probes for CP-violating effects beyond the Standard Model.
The upper limits obtained in various precision experiments can therefore be used to strongly restrict new physics models. 
However, relating the experimental information to parameters of a specific model is complicated by the presence of various sources for EDMs as well as large theory uncertainties in some of the relevant matrix elements.
In this article, we address both issues for the EDMs of heavy paramagnetic systems, where it is possible to include subleading contributions, thereby model-independently extracting the electron EDM. We furthermore use expressions for the presently phenomenologically relevant EDMs with conservative estimates for the theoretical uncertainties to place constraints on CP-violating phases in the context of Two-Higgs-Doublet models.
}

\section{Introduction}
Electric dipole moments (EDMs) provide a competitive means to search for new physics (NP), complementary to strategies like direct searches at hadron colliders, but also to other indirect searches like the flavour-changing processes investigated at the flavour factories. The exceptional sensitivity is in two ways related to the very specific connection between flavour and CP violation\footnote{EDMs are T,P-odd, implying also CP violation when assuming CPT to conserved as we will in this article.} in the SM, embodied by the Kobayashi-Maskawa mechanism:\cite{Kobayashi} firstly, it is very effective in suppressing flavour-changing neutral currents (FCNCs), and even more so flavour-conserving ones involving CP violation. An exception is provided by the gluonic operator $\mathcal{O}_{G\tilde G}\propto\epsilon_{\mu\nu\rho\sigma}G^{\mu\nu}G^{\rho\sigma}$: its potentially very large contribution to hadronic EDMs is, however, strongly bounded experimentally. 
In this work it is implicitly assumed that it is effectively removed by Peccei-Quinn symmetry\cite{Peccei:1977hh} or a similar mechanism. The remaining SM contributions then lead to EDMs many orders of magnitude below the present limits, \emph{e.g.}\cite{Khriplovich:1981ca,Gavela:1981sk,McKellar:1987tf,Mannel:2012qk} $d_n^{\rm SM,CKM}\lesssim(10^{-32}-10^{-31})\,e\,{\rm cm}$. Importantly, for leptonic EDMs no assumption regarding $\mathcal{O}_{G\tilde G}$ is necessary; the SM contribution to the electron EDM is estimated to be\cite{Pospelov:1991zt,Booth:1993af,Pospelov:2013sca} $d_e^{\rm SM}\lesssim10^{-38}\,e\,{\rm cm}$.
The observation of an EDM with the present experimental precision would therefore clearly constitute a NP signal, especially in the leptonic sector. The second way the Kobayashi-Maskawa mechanism plays a role is that in a generic NP scenario, the absence of such a powerful suppression typically yields contributions that are large compared to experimental limits. On the other hand, Sakharov's conditions \cite{Sakharov:1967dj} require the presence of new sources of CP violation to explain the observed baryon asymmetry in the universe; while this does not necessarily imply sizable EDMs, it yields a strong motivation to search for such sources.
This combination of tiny SM ``background'' and comparatively large expected NP contributions renders EDMs a precision laboratory for NP searches. 

A potential discovery of a non-vanishing EDM would therefore indicate a NP signal several orders of magnitude above the SM expectation, rather independent of theoretical uncertainties or the specific source. However, when casting existing experimental limits (and also potential signals) into bounds on model parameters, both issues need to be addressed. Specifically, since experiments are typically carried out using composite systems, that is, nucleons, atoms or molecules, the relation to more fundamental quantities like the electron EDM requires the evaluation of complicated matrix elements which constitute the main source of theoretical uncertainty. Furthermore, there are different sources for EDMs in theories beyond the SM, which can exhibit cancellations. For heavy paramagnetic systems, potential cancellations can be taken into account, leading to a more reliable, model-independent limit on the electron EDM,\cite{Jung:2013mg} discussed in the next section.
This is then used together with other limits in section~\ref{sec::2HDMs} to constrain the CP-violating parameters in Two-Higgs-Doublet models (2HDMs). Specifically, models with new CP-violating phases in the Yukawa interactions used to be discarded because of potentially huge EDMs. While the present experimental limits impose strong bounds on the corresponding parameters, we show that in models with an appropriate flavour structure they have not yet to be unnaturally small.\cite{Jung:2013hka} However, large enhancements in other CP-violating observables are strongly restricted by these bounds.
Furthermore, the generic size for EDMs lies well within reach of planned and ongoing next-generation experiments. These will therefore provide critical tests for this class of models.
We conclude in section~\ref{sec::conclusions}.

\section{Framework}
Relating experimental data to fundamental parameters proceeds in a series of effective theories. The available competitive observables, that is, the EDMs of thorium monoxide and ytterbium fluoride  molecules\cite{Baron:2013eja,Hudson:2011zz}, thallium and mercury atoms\cite{Regan:2002ta,Griffith:2009zz} and the neutron\cite{Baker:2006ts} (see also \cite{Serebrov:2013tba}), are related by atomic, nuclear and QCD calculations to the coefficients of an effective theory on a hadronic scale (see, \emph{e.g.}, Ref.\citelow{Pospelov:2005pr}):
\begin{eqnarray}\label{eq::Leff}
\mathcal{L}^{\rm EDM}_{\rm eff} = -\!\!\sum_{f}\frac{d_f^\gamma}{2}\mathcal{O}_f^\gamma-\sum_q\frac{d_q^C}{2}\mathcal{O}_q^C+C_W \mathcal{O}_W+\sum_{f,f'}C_{ff'}\mathcal{O}_{ff'}\,.
\end{eqnarray}
The operator basis consists of (colour-)EDM operators $\mathcal{O}^{\gamma,C}_f$ ($f=e,q$, $q=u,d,s$), the Weinberg operator $\mathcal{O}_W$ and T- and P-violating four-fermion operators $\mathcal{O}_{ff'}$ without derivatives (see,\emph{e.g.}, Ref.\citelow{Khriplovich:1997ga}).
Since these calculations do not depend on the NP model under consideration, this is used as the interface between the experimental side and the high-energy calculations: the latter provide the model-specific expressions for the Wilson coefficients in Eq.~\eqref{eq::Leff}, with at least one more intermediate effective theory at the electroweak scale.

\section{Model-independent extraction of the electron EDM\label{sec::eEDMMI}}
\begin{figure}[hbt]
\centering{
\includegraphics[width=0.33\textwidth]{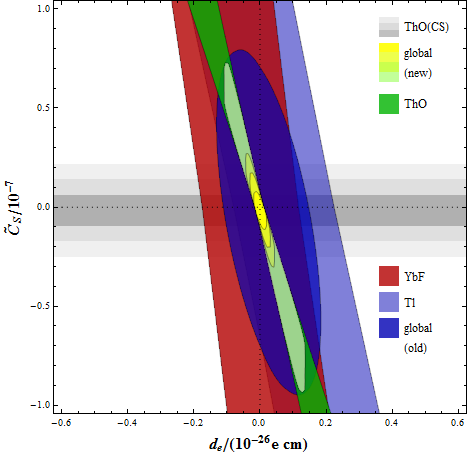}\quad\includegraphics[width=0.33\textwidth]{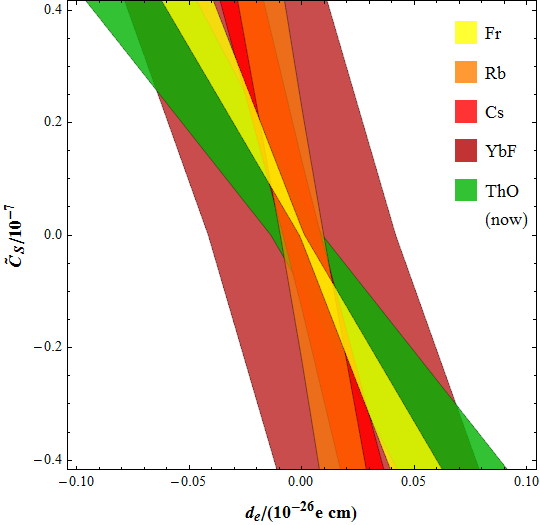}\quad
\begin{minipage}[b]{0.28\textwidth}
\caption{\label{fig::eEDM}
The constraint for the electron EDM ($95\%$~CL) from the measurements in paramagnetic systems, see text. Left: global fit in comparison to the results before the $ThO$ measurement. Right: comparison of the $ThO$ measurement the projected results of ongoing experiments.\hfill\mbox{}\newline} 
\end{minipage}
}
\end{figure}

Within a given model, typically different operators from Eq.~\eqref{eq::Leff} dominate in different regions of the parameter space. Heavy paramagnetic systems are an exception in this respect: their EDMs receive two contributions scaling at least like $d\sim Z^3$,\cite{Sandars:1965xx,Sandars:1966xx,Flambaum:1976vg} which therefore dominate the others; one term is directly proportional to the electron EDM $d_e$, the other stems from electron-nucleon interactions, parametrized by a dimensionless parameter $\tilde C_S$. The energy shift $\Delta E=\hbar\omega$ of molecules $M$ in an external field, as measured in,\cite{Baron:2013eja,Hudson:2011zz} is given in terms of these contributions as well:
\begin{equation}
\omega = 2\pi \left(\frac{W_d^M}{2} d_e+\frac{W_c^M}{2}\tilde C_S\right)\,.
\end{equation}
The necessary constants $W_{d,c}^M$, as estimated in \cite{Jung:2013mg,Jung:2013hka}, are $W_d^{YbF}=-(1.3\pm0.1)10^{25}\,{\rm Hz}/e~{\rm cm}$, $W_c^{YbF}=-(92\pm9)\,{\rm kHz}$,  $W_d^{ThO}=-(3.67\pm0.18)10^{25}\,{\rm Hz}/e~{\rm cm}$, $W_c^{ThO}=-(598\pm90)\,{\rm kHz}$. In the literature it is common, however, to extract the electron EDM by setting $\tilde C_S\to0$ (and neglecting theory uncertainties). While this is reasonable in some models, it is not a model-independent procedure. Actually, a single measurement cannot be translated into a limit on $d_e$ without an assumption on $\tilde C_S$. In principle, however, clearly both contributions can be extracted, once more than one measurement is available. Using additionally the measurement for mercury, this has been done in Ref.\cite{Jung:2013mg}, leading to a bound on $d_e$ competitive with the naive extraction at the time. 
The present situation, illustrated in Fig.~\ref{fig::eEDM}\cite{Jung:2013mg,Jung:2013hka} on the left, is that the in principle much stronger limit\cite{Baron:2013eja} cannot easily be translated into a much better constraint on the electron EDM (comparing the projections on the $d_e$ axis of the blue ellipsis versus the one of its overlap with the dark green fan yields and improvement from $|d_e|\leq 1.4\times 10^{-27}\,e~{\rm cm}$ to $|d_e|\leq 1.0\times 10^{-27}\,e~{\rm cm}$, only), since no second competitive measurement is available. 
This will change in the future, as illustrated in the same figure on the right. Note that it is of special importance to have measurements with significantly differing values of $W_d^M/W_c^M$,\cite{Jung:2013mg} that is, measurements with atoms/molecules of different weight.\cite{PhysRevA.85.029901}
In the meantime we propose to use a fine-tuning argument instead of neglecting the $\tilde C_S$ contribution completely: allowing the latter by itself to saturate the experimental limit at most $n=1,2,3,\ldots$ times, \emph{i.e.} excluding very large cancellations, we obtain a still conservative upper limit on $d_e$, \emph{e.g.}\cite{Jung:2013hka}
\begin{equation}
|d_e|\leq 0.25\times 10^{-27}e\,{\rm cm}\quad (n=2)\,.
\end{equation}

\section{EDMs in Two-Higgs-Doublet Models\label{sec::2HDMs}}
We calculate the Wilson coefficients in Eq.~\eqref{eq::Leff} for 2HDMs with new CP-violating phases. To that aim, we use a general parametrization for the charged current Yukawa couplings in the Higgs basis,
\begin{equation}\label{eq::LHcharged}
\mathcal L_Y^{H^\pm} \! =\! - \frac{\sqrt{2}}{v}\, H^+ \left\{  \bar{u}   \left[  V \varsigma_d M_d \mathcal P_R - \varsigma_u\, M_u^\dagger V \mathcal P_L  \right]   d\, +\,  \bar{\nu} \varsigma_l M_l \mathcal P_R l\right\} + \;\mathrm{h.c.} \, ,
\end{equation}
where the $M_i$ are diagonal mass matrices, $V$ denotes the CKM matrix, and the $\varsigma_f$ in principle arbitrary complex matrices. We give below the constraints in terms of elements of these matrices, which can be translated into the parameters of any given 2HDM model. To be specific and able to relate the resulting bounds also to those from other observables, we will furthermore consider the Aligned 2HDM (A2HDM),\cite{Pich:2009sp,Jung:2010ik} where the $\varsigma_i$ are complex numbers, thereby avoiding FCNCs on tree level while still allowing for a rich phenomenology including additional CP-violating phases. 
For the couplings of the neutral Higgs states, we obtain similarly
\begin{equation}\label{eq::LHneutral}
\mathcal L_Y^{\varphi_i^0} \! =\! -\frac{1}{v}\; \sum_{\varphi, f}\,  \varphi^0_i  \; \bar{f}\,y^{\varphi^0_i}_f\,  M_f \mathcal P_R  f\;  + \;\mathrm{h.c.} \, ,
\end{equation}
with the fields $\varphi_i^0=\{h,H,A\}$ denoting the neutral scalar mass eigenstates. Denoting the fermion species as $F(f)$,  
\emph{e.g.} $F(u)=F(c)=F(t)=u$, 
we can write the appearing couplings as \emph{e.g.} $y_{f}^{\varphi^0_i} = \cR_{i1} + (\cR_{i2} + i\,\cR_{i3})\,\left(\varsigma_{F(f)}\right)_{ff}$  (for $F(f)=d,l$) to allow for the general form of $\varsigma_{u,d,l}$.
$\cR$ denotes the rotation defined by $\mathcal{M}^2_{\rm diag}=\mathcal{R} \mathcal{M}^2\mathcal{R}^T$, relating the mass eigenstates to the neutral scalar fields in the Higgs basis. In a general 2HDM, the  $\varsigma_{u,d,l}$ are the matrices introduced in Eq.~\eqref{eq::LHcharged}, only the diagonal elements of which are relevant here.
This expression for the couplings $y_{f}$ reflects the fact that in the neutral Higgs couplings CP violation may enter from the Yukawa couplings as well as the scalar potential, rendering the phenomenological discussion very complicated. However, orthogonality of the matrix $\cR$ implies the relation
\begin{equation}\label{eq::ycancellation}
\sum_i {\rm Re}\left(y_f^{\varphi^0_i}\right){\rm Im}\left(y_{f'}^{\varphi^0_i}\right)=\pm{\rm Im}\left[(\varsigma_{F(f)}^*)_{ff}(\varsigma_{F(f')})_{f'f'}\right]\,,
\end{equation}
which vanishes for real $\varsigma_i$ (as \emph{e.g.} the case for $\mathcal Z_2$ models) and for $f=f'$ (fermions of the same family if the $\varsigma_i$ are family-universal as \emph{e.g.} in the A2HDM). Importantly, the right-hand side is independent of the parameters of the scalar potential. While in practical calculations there are mass-dependent weight factors in the sum on the left, the relation still holds exactly in two limits: trivially so when the neutral scalars are degenerate, but also in the decoupling limit.\cite{Jung:2013hka} Therefore, in general cancellations can be expected for any mass spectrum and the influence of CP violation in the potential is reduced. Clearly, this observation provides a protection against large EDMs for models which exhibit new CP-violating parameters in the potential, only. Below we will assume this relation to hold and evaluate the right-hand side with a common weight factor at an intermediate effective neutral Higgs mass $\overline{M}_\varphi$.

The fact that so far no significant NP signals have been observed directly implies a highly non-trivial flavour structure of the theory. In models fulfilling this requirement,  the main contributions then stem typically from two-loop diagrams, namely the Weinberg operator and so-called Barr-Zee diagrams\cite{Weinberg:1989dx,Barr:1990vd}; additionally, enhanced four-fermion operators can be important for heavy atoms/molecules, as emphasized above. We refer the reader to\cite{Jung:2013hka} for the relevant expressions in 2HDMs and show below directly the resulting constraints.

\begin{figure}[htb]
\centering{
\includegraphics[width=0.33\textwidth]{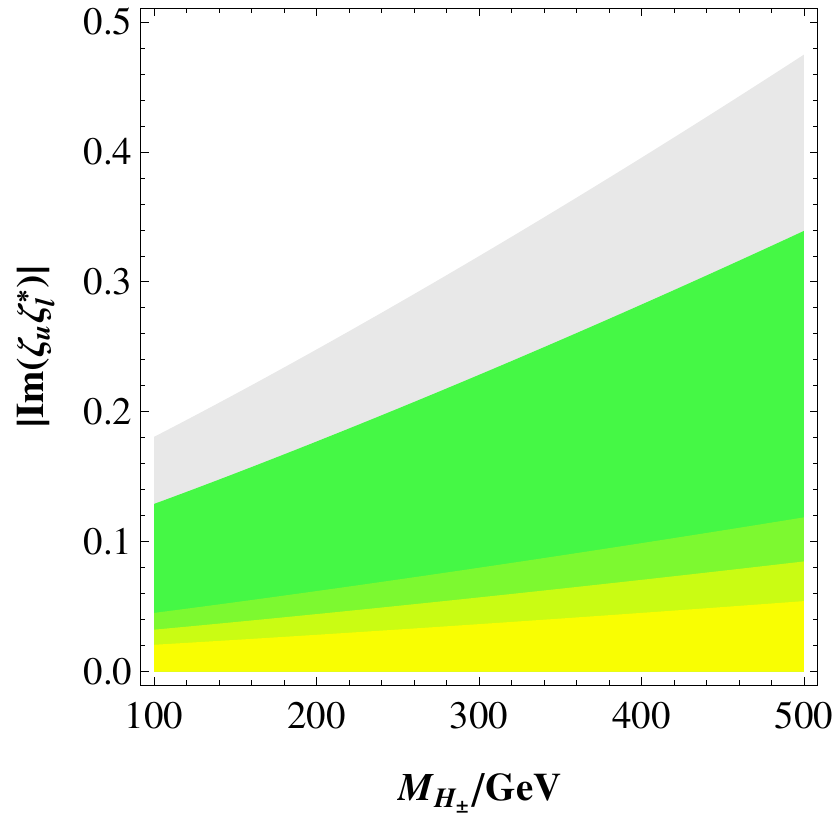}\quad\includegraphics[width=0.33\textwidth]{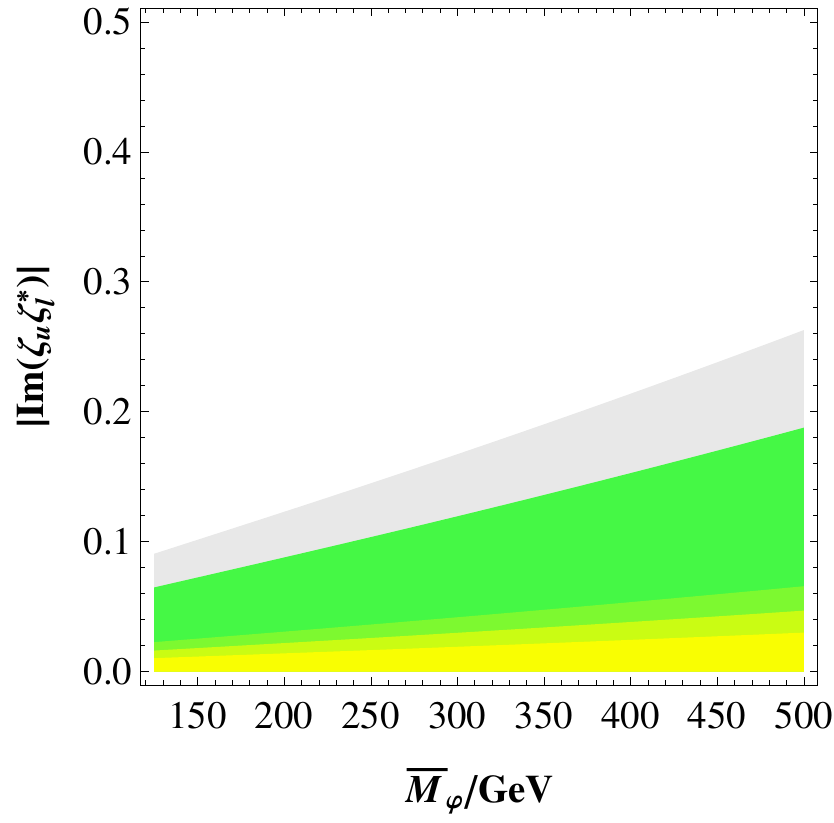}
\quad
\begin{minipage}[b]{0.28\textwidth}
\caption{\label{fig::A2HDMeEDM}Constraints from the electron EDM ($95\%$~CL) on charged/neutral Higgs exchange (left/right) in the ${\rm Im}(\varsigma_u^*\varsigma_l)-M_{H^\pm}/\overline M_{\varphi}$ plane. In grey is shown the result without the $ThO$ measurement, dark green the extremely conservative model-independent bound, while the other areas correspond to $n=1,2,3$, cf. Sec.~\ref{sec::eEDMMI}.\hfill\mbox{}\newline}
\end{minipage}
}
\end{figure}

The electron EDM receives contributions mostly from Barr-Zee diagrams.\cite{BowserChao:1997bb} The resulting constraints on ${\rm Im}(\varsigma_{u,33}\varsigma_{l,11}^*)$ are shown in Fig.~\ref{fig::A2HDMeEDM}, demonstrating the strength of this observable. 
For the A2HDM, this becomes even more obvious when comparing with the bound on the absolute value of this parameter combination obtained from leptonic and semileptonic decays,\cite{Jung:2010ik,Celis:2012dk} which is about a factor 1000 weaker.

\begin{figure}[htb]
\centering{
\includegraphics[width=0.33\textwidth]{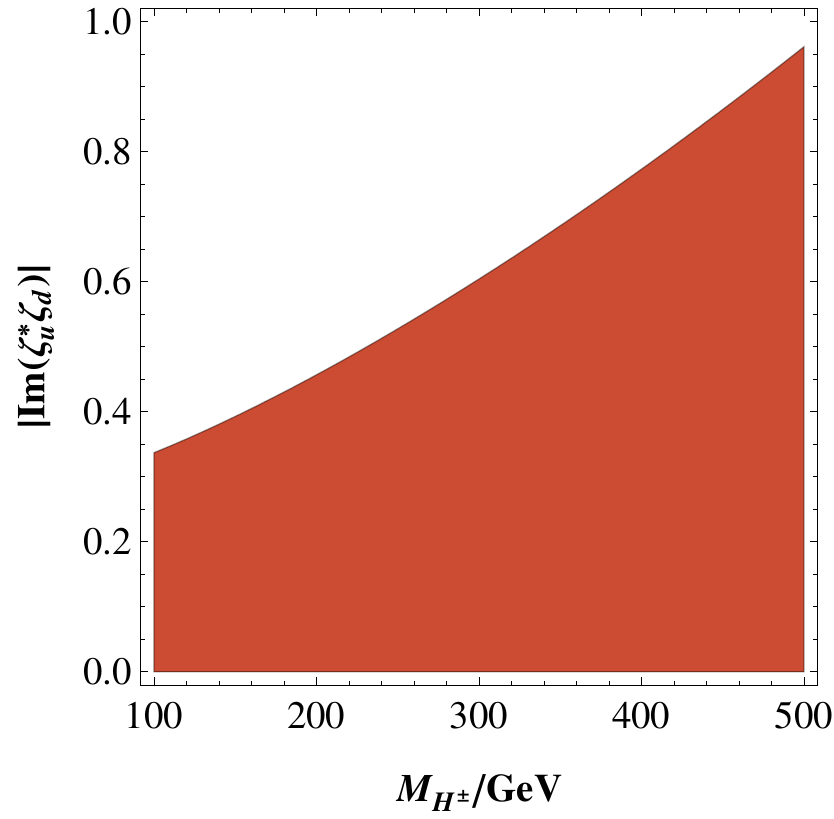}\quad\includegraphics[width=0.33\textwidth]{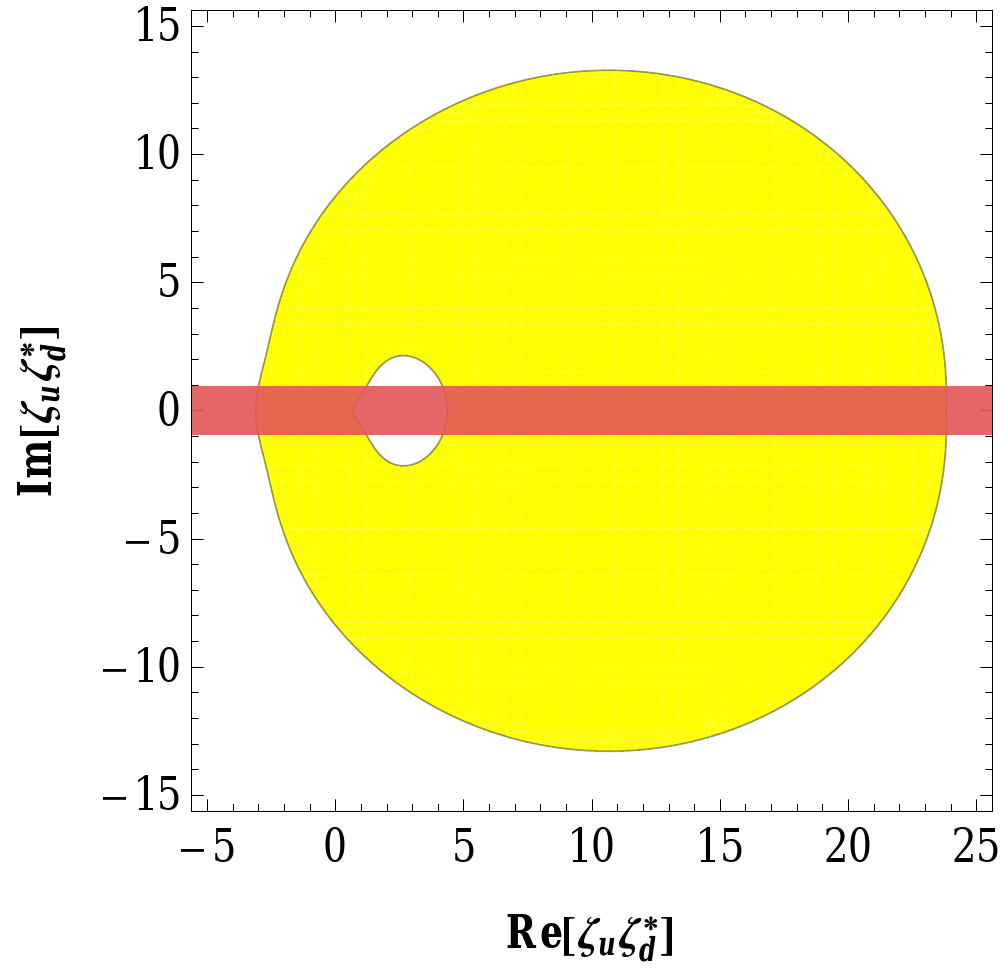}
\quad
\begin{minipage}[b]{0.28\textwidth}
\caption{\label{fig::nEDMWzetaud}The constraint from the neutron EDM ($95\%$~CL) in the ${\rm Im}(\varsigma_{u,33}^*\varsigma_{d,33})-M_{H^\pm}$ plane (left) and in the A2HDM together with the constraint from $BR(b\to s\gamma)$  in the complex $\varsigma_u\varsigma_d^*$ plane (right).\hfill\mbox{}\newline}
\end{minipage}}
\end{figure}

For the neutron, the constraint induced in the charged-Higgs sector via the Weinberg operator is shown in Fig.~\ref{fig::nEDMWzetaud} on the left. 
Again no fine-tuning is necessary to avoid this bound. On the other hand it prohibits large CP-violating effects in other observables. On the right, the maximally allowed band is shown together with the constraint from the branching ratio in $b\to s\gamma$, again for the A2HDM; from the discussion in Refs.\cite{Jung:2010ab,Jung:2012vu} follows for the NP contribution that $|A_{\rm CP}(b\to s\gamma)|\lesssim1\%$.

\section{Conclusions\label{sec::conclusions}}
EDMs provide unique constraints for the CP-violating sectors of NP models. We discussed the model-independent extraction of the electron EDM from measurements in heavy paramagnetic systems and the application of EDM constraints to general 2HDMs. While so far no severe fine-tuning is necessary to avoid the resulting bounds, they prohibit large effects in other CP-violating observables in concrete models like the A2HDM. Given the present strength of the constraints, forthcoming experiments will test a crucial part of the parameter space and might turn existing bounds into observations.

\section*{Acknowledgments}
I would like to thank Toni Pich for a fruitful and enjoyable collaboration, as well as the organizers and participants of Moriond EW 2014 for a very pleasant conference. This work is supported in part by the Bundesministerium f\"ur Bildung und Forschung (BMBF).

\section*{References}
\bibliography{EDM_proceedings}

\end{document}